\documentclass[aps,prd,12pt,nofootinbib]{revtex4}
\usepackage{epsfig}
\usepackage{graphicx}
\usepackage[font=small,skip=0pt,justification=raggedright]{caption}
\usepackage{amsmath}
\usepackage{amssymb}
\usepackage{mathrsfs}
\usepackage{verbatim}
\usepackage{dutchcal}

\usepackage[normalem]{ulem}
\usepackage{xcolor}

\newcounter{fig}

\newcommand{\bea}{\begin{eqnarray}}
\newcommand{\eea}{\end{eqnarray}}
\newcommand{\be}{\begin{equation}}
\newcommand{\ee}{\end{equation}}

\newcommand{\re}[1]{(\ref{#1})}

\newcommand{\eqn}{\begin{eqnarray}}
\newcommand{\eqnx}{\end{eqnarray}}

\begin{document}

\title{Boson Stars}
\author{Yakov Shnir}
\affiliation{BLTP, JINR, Dubna 141980, Moscow Region, Russia}

%
\begin{abstract}
We review particle-like configurations of complex scalar field,
localized by gravity, so-called boson stars. In the simplest case,
these solutions posses spherical symmetry, they may arise in the
massive Einstein-Klein-Gordon theory with global $U(1)$ symmetry,
as gravitationally bounded lumps of scalar condensate. Further,
there are spinning axially symmetric boson stars which possess
non-zero angular momentum, and a variety of non-trivial multipolar
stationary configurations without any continuous symmetries. In
this short overview we discuss important dynamic properties of the
boson stars, concentrating on recent results on the construction
of multicomponent constellations of boson stars.
\end{abstract}

\maketitle

\section{Q-balls and Boson Stars}
\label{sec:1}

One of the most interesting directions in modern theoretical physics is related with investigation of spacially localized
field configurations with finite energy bounded by gravity, see i.e.
\cite{Jetzer:1991jr,Liddle:1992fmk,Schunck:2003kk,Liebling:2012fv} for detailed review. The idea that the
gravitational attraction may stabilize a fundamental matter field, was pioneered by Wheeler \cite{Wheeler:1955zz},
who considered classical self-gravitating lumps of electromagnetic field, so-called \emph{geons}. The geon is a
localized regular solution of the coupled system of the field equations of the Einstein-Maxwell theory. Notably, Wheeler emphasized
the unstability of geons with respect to linear perturbations of the fields.

From a modern perspective, the geons represent a \emph{soliton}, a field configuration
which may exist in diverse non-linear models in a wide variety of
physical contexts. Roughly speaking, the solitons can be can be divided into two groups,
the topological and non-topological solitons, see e.g. \cite{Manton:2004tk,Shnir2018}.
Topological solitons, like kinks, vortices, monopoles or skyrmions,
are characterized by  a conserved topological charge. This is not a property of non-topological solitons which occur in various non-linear
systems with an unbroken global symmetry. A typical example in
Minkovski spacetime are Q-balls, they represent time-dependent lumps of a complex scalar field with a stationary oscillating  phase
\cite{Rosen,Friedberg:1976me,Coleman:1985ki}.

It was pointed out by  Kaup \cite{Kaup:1968zz}, Feinblum and McKinley \cite{Feinblum:1968nwc},
and subsequently by Ruffini and Bonazzola \cite{Ruffini:1969qy},
that stable localized soliton-type configurations, now dubbed as \emph{boson stars} (BSs) may arise
as the complex scalar field becomes coupled to gravity. In the simplest case, spherically symmetric boson star represent
a particle-like self-gravitating asymptotically flat stationary solution of the (3+1)-dimensional
Einstein-Klein-Gordon (EKG) theory. In this model the scalar field possess a mass term only, without self-interaction.
The corresponding configurations can be considered as lumps of the scalar condensate,
macroscopic quantum state, which is prevented from
gravitationally collapsing by Heisenberg$^\prime$s uncertainty principle.
These mini-boson stars do not have a
regular flat spacetime limit. On the contrary, the BSs in the models
with polynomial potentials \cite{Kleihaus:2005me,Kleihaus:2007vk}, or in the two-component Einstein-Friedberg-Lee-Sirlin model
\cite{Friedberg:1986tq}, are linked to the corresponding flat space Q-balls.
The BSs in the model with a repulsive self-interaction \cite{Colpi:1986ye} are more massive than the mini-boson stars in the EKG model,
further, inclusion of a sextic potential \cite{Friedberg:1986tq,Kleihaus:2005me,Kleihaus:2007vk} allows for existence of very massive and highly compact
objects, near of the threshold of gravitational collapse \cite{Lee:1986ts,Hawley:2000dt}. Clearly, these configurations resemble neutron stars, further
astrophysical applications of BSs include consideration of hypothetical weakly-interacting
ultralight component of cosmological dark matter \cite{Suarez:2013iw,Hui:2016ltb}, axions \cite{Guerra:2019srj,Delgado:2020udb},
and black hole mimickers \cite{Cardoso:2019rvt,Glampedakis:2017cgd,Herdeiro:2021lwl}.
Bosons stars attracted a lot of attention in study of their evolution in binaries and in search
for gravitational-wave signals produced by collision of boson stars \cite{Palenzuela:2007dm,Bezares:2017mzk,Palenzuela:2017kcg}.

Both Q-balls and BSs have a harmonic time dependence with a constant angular frequency $\omega$, they
carry a Noether charge $Q$ associated with an unbroken continuous global $U(1)$ symmetry. This
charge is proportional to the frequency $\omega$ and represents the boson particle number of the configurations.
Further, there are charged Q-balls in  gauged models with local $U(1)$ symmetry
\cite{Lee:1988ag,Lee:1991bn,Kusenko:1997vi,Anagnostopoulos:2001dh,Gulamov:2015fya,Gulamov:2013cra,Panin:2016ooo,Nugaev:2019vru,Loiko:2019gwk}.
The presence of the electromagnetic interaction affects the properties of the gauged Q-balls, in particular, they may exist for
a restricted range of values of the gauge coupling.
Charged BSs arise in extended Einstein-Maxwell-scalar theories, these solutions were
studied in \cite{Jetzer:1989av,Jetzer:1989us,Jetzer:1992tog,Pugliese:2013gsa,Kleihaus:2009kr,Kumar:2014kna}.
Besides, BSs exist in the asymptotically anti-de Sitter spacetime \cite{Astefanesei:2003qy,Kichakova:2013sza}.

In  Minkowski  spacetime, Q-balls exist only within a restricted interval of values of the angular frequency $\omega$:
there is a maximal value $\omega_{max}$, which corresponds to the mass of the scalar
excitations, and some minimal value $\omega_{min}$, that depends on the form of the potential. Notably, $\omega_{min}=0$ in the
two-component Friedberg-Lee-Sirlin (FLS) model \cite{Friedberg:1976me,Levin:2010gp,Loiko:2018mhb}.
Both the mass $M$ and the charge $Q$ diverge, as the frequency $\omega$ approaches the limiting values.
Typically, there are two branches of flat space Q-balls, merging and
ending at the minimal values of charge and mass. This bifurcation corresponds to some critical value of the
frequency $\omega_{cr} \in [\omega_{min}, \omega_{max}]$, from where they increase monotonically towards both
limiting values of $\omega$.

The situation is different for BSs: coupling of the scalar field to gravity modifies the
critical behavior pattern of the configurations. The fundamental
branch of the solutions starts off from the perturbative excitations at $\omega \sim \omega_{max}$, at which both
the mass and the charge trivialize (rather than diverge). Then, the BSs exhibit
a spiral-like frequency dependence of the charge and the mass, where both quantities tend to some
finite limiting values at the centers of the corresponding spirals \cite{Friedberg:1986tq}.
Qualitatively, the appearance of the frequency-mass spiral
may be related to oscillations in the force balance between
the repulsive scalar interaction and the gravitational attraction
in equilibria \cite{Friedberg:1986tp}.
This spiraling behavior is reminiscent of the mass radius relation of neutron
stars beyond the maximum mass star.

Simplest BSs are spherically symmetric, for each fundamental solutions there exist a tower of
radially excited states, which possess some number of nodes in profile of the scalar
field \cite{Seidel:1990jh,Jetzer:1991jr,Friedberg:1986tq}.
The mass of these excited solution is higher, than the mass of the corresponding fundamental boson star with the
same angular frequency $\omega$, however the properties of the spherically symmetric
excited BSs are not very different from
those of the nodeless boson stars. Also multi-state BSs have been studied, these configurations
represent spherically symmetric superposition of the fundamental and the first excited solutions \cite{Bernal:2009zy}.
The radial pulsations and radiation of BSs were studied in numerical
relativity \cite{Hawley:2000dt,Gleiser:1988ih,Kain:2021rmk}, the solutions are shown to be stable on the first branch.

Rotating BSs are axially symmetric, they possess non-zero angular momentum $J$ which is quantized in terms of
the charge, $J = n Q$ \cite{Silveira:1995dh,Schunck:1996he,Kleihaus:2005me,Kleihaus:2007vk}.
In other words, the BSs do not
admit slow rotating limit. Rotating BSs possess some
peculiar geometrical features, in particular, ergo-regions may arise for such solutions \cite{Kleihaus:2007vk,Cardoso:2007az}.
Interestingly, radially excited rotating BSs do not exhibit a spiraling behavior; instead, the second branch
extends back to the upper critical value of the
frequency $\omega_{max}$, forming a loop \cite{Collodel:2017biu}.

Both axially-symmetric spinning Q-balls in  Minkowski spacetime  and the rotating BSs
may be either symmetric with respect to reflections in the equatorial
plane, $\theta \to \pi -\theta$, or antisymmetric.
The solutions of the first type are referred to as parity-even,
while the configurations of the second type are termed parity-odd
\cite{Yoshida:1997qf,Volkov:2002aj,Kleihaus:2005me,Kleihaus:2007vk,Radu:2008pp},
for each value of integer winding number $n$,
there should be two types of spinning solutions possessing
different parity.

Notably, the character of the scalar interaction
between Q-balls and BSs depends on their relative phase \cite{Battye:2000qj,Bowcock:2008dn},
If the solitons are in phase, the scalar interaction is attractive, if they are out of phase, there is a
repulsive scalar force between them. Thus, a pair of boson stars may exist as a saddle point solution
of the EKG model \cite{Yoshida:1997nd,Herdeiro:2021mol,Herdeiro:2020kvf}. Furthermore, scalar repulsion can be
balanced by the gravitational attraction in various multicomponent bounded systems of BSs \cite{Herdeiro:2021mol,Herdeiro:2020kvf}.

Below we briefly review the basic properties of boson stars and discuss multicomponent BS configurations
constructed recently in \cite{Herdeiro:2021mol,Herdeiro:2020kvf}.

\section{The model: Action, field equations, and global charges}
We consider a massive complex scalar field $\Phi$,
which is minimally
coupled to Einstein's gravity in an asymptotically flat
$(3+1)$-dimensional space-time. The corresponding action of
the system is
\begin{equation}
\label{action}
\mathcal{S}=\int  d^4x \sqrt{-g}\left[ \frac{R}{16\pi G}
   -\frac{1}{2} g^{\mu\nu}\left( \Phi_{, \, \mu}^* \Phi_{, \, \nu} + \Phi _{, \, \nu}^* \Phi _{, \, \mu} \right) - U(|\Phi|^2 )
 \right] ,
\end{equation}
where $R$ is the Ricci scalar curvature, $G$ is Newton's constant,
the asterisk denotes complex conjugation,
$U $ denotes the scalar field potential
and we employ the
usual compact notation $\Phi_{, \, \mu} \equiv \partial_\mu \Phi $.

Variation of the action \re{action} with respect to the metric leads to the Einstein equations
\be
\label{Einstein}
E_{\mu\nu}\equiv R_{\mu\nu}-\frac{1}{2}g_{\mu\nu}R-8 \pi G~T_{\mu\nu}=0 \ ,
\ee
where
\be
\label{SET}
T_{\mu\nu}\equiv
 \Phi_{ , \mu}^*\Phi_{,\nu}
+\Phi_{,\nu}^*\Phi_{,\mu}
-g_{\mu\nu}  \left[ \frac{1}{2} g^{\sigma\tau}
 ( \Phi_{,\sigma}^*\Phi_{,\tau}+
\Phi_{,\tau}^*\Phi_{,\sigma} )+U(|\Phi|^2) \right] \, ,
\ee
is the stress-energy tensor of the scalar field.

The corresponding equation of motion of the scalar field
is the non-linear Klein-Gordon equation
\be
\label{scaleq}
    \left(\Box - \frac{d U}{d |\Phi|^2} \right)\Phi=0\, ,
\ee
where $\Box$ represents the covariant d'Alembert operator.

The solutions considered below have a static line-element
(with a  timelike Killing vector field  $\xi=\partial_t$),
  being
topologically trivial and
  globally regular, $i.e.$ without an event horizon or conical singularities,
while the scalar field is finite and smooth everywhere.
Also, they approach asymptotically the Minkowski spacetime background.
Their mass $M$
can be obtained from the respective Komar expressions \cite{wald},
\begin{equation}
\label{komarM}
{M} = 2 \int_{\Sigma}
 R_{\mu\nu}n^\mu\xi^\nu dV~
 = \,  2 \int_{\Sigma} \left(  T_{\mu \nu}
-\frac{1}{2} \, g_{\mu\nu} \, T_{\gamma}^{\ \gamma}
 \right) n^{\mu }\xi^{\nu} dV .
\end{equation}
Here $\Sigma$ denotes a  spacelike hypersurface
(with  the  volume element $dV$),
while
$n^\mu$ is a time-like vector normal to $\Sigma$,  $n_\mu n^\mu = -1$.

The axially symmetric spinning boson stars are characterized by the mass $M$ and by
the angular momentum
\be
\label{komarJ}
J =  -\int_{\Sigma}
 R_{\mu\nu}n^\mu\eta^\nu dV = - \int_{\Sigma}
 \left(  T_{\mu \nu}
-\frac{1}{2} \, g_{\mu\nu} \, T_{\gamma}^{\ \gamma}
 \right) n^{\mu }\eta^{\nu} dV
\ee
where the second commuting Killing vector field is $\eta=\partial_\varphi$.

The action \re{action} is invariant with respect to the global $\mathrm{U}(1)$
transformations of the complex scalar
field, $\phi\to\phi e^{i\chi }$, where $\chi$ is a constant.
The following Noether 4-current is associated
with this symmetry
\be
\label{Noether}
j_\mu =- i(\Phi\partial_\mu\Phi^\ast-\Phi^\ast\partial_\mu\Phi)\, .
\ee
It follows that integrating the timelike component of this 4-current in a spacelike slice
$\Sigma$ yields a second conserved quantity -- the \textit{Noether charge}:
\begin{eqnarray}
\label{Q}
Q =\int_{\Sigma} j^{\mu}n_\mu dV\, .
\end{eqnarray}

Semiclassically, the charge $Q$ can be interpreted as a measure of the number of scalar quanta
condensed in the BS. There is the quantization
relation for the angular momentum of the scalar field $J$ \re{komarJ}, $J=nQ$ \cite{Schunck:1996he}.

\subsection{Potential}

In the simplest case of the non-self interacting EKG model, the potential
contains just the mass term, $U= \mu^2|\Phi|^2$,
where parameter  $\mu$ yields the mass of the scalar field. The corresponding mini-BSs
represent a gravitationally bound system of globally regular massive interacting bosons, it
does not possess the flat space limit. It should be noted that in the EKG model the natural units are set by
the mass parameter $\mu$ and by the effective gravitational coupling $\alpha^2=4\pi G$.
They can be rescaled away via transformations of
the coordinates and the field, $x_\mu\to x_\mu/\mu$, $\Phi \to \Phi/\alpha$. Note that the
scalar field frequency changes accordingly, $\omega \to \omega/\mu$.

The quartic self-interaction potential
\be
U = \lambda |\Phi|^4 + \mu^2 |\Phi|^2 \, ,
\label{pot-quart}
\ee
was considered in many works, see e.g. \cite{Colpi:1986ye,Herdeiro:2015tia,Sanchis-Gual:2021phr}. Such potential can
stabilize excited BSs, however the corresponding solutions do not posses the flat space limit.

The non-renormalizable self-interacting sixtic potential, originally proposed in \cite{Deppert:1979au,Mielke:1980sa}
\be
U = \nu |\Phi|^6- \lambda |\Phi|^4 + \mu^2 |\Phi|^2
\label{pot-six}
\ee
allows for the existence of very massive BSs, they are linked to the corresponding Q-balls on a Minkowski spacetime
background \cite{Friedberg:1986tq,Volkov:2002aj,Kleihaus:2005me,Radu:2008pp}. Similar to the case of the EKG model,
two of the parameters of the model \re{action},\re{pot-six} can be absorbed
into a redefinition of the coordinates together with a rescaling of the scalar field,
$$
x_\mu\to \frac{a}{\mu} x_\mu \, , \quad \Phi \to \frac{\sqrt \mu}{\nu^{1/4} \sqrt a} \Phi \, ,
$$
where $a$ is an arbitrary constant. Thus, the potential of the rescaled model becomes
$$
U=|\Phi|^6 - \tilde \lambda |\Phi|^4 + a^2 |\Phi|^2
$$
with the usual choice $\tilde \lambda =\frac{a\lambda}{\mu \sqrt \nu}=2$ and $a^2=1.1$. Then the
dimensionless effective gravitational coupling becomes
$\alpha^2=\frac{4\pi G \mu}{a \sqrt \nu}$. Evidently, for large values of the
gravitational coupling, the nonlinearity of the
potential  \re{pot-six} becomes suppressed and the system
approaches the EKG model with its corresponding mBS solutions. However, as the gravitational attraction remains
relatively weak, the scalar interaction becomes more important, it allows for existence of very large massive BSs.

The sixtic potential \re{pot-six} can be considered as a limiting form of the periodic  axion potential which
describes a real quantized scalar field $\Phi$,
$$
U=m_a f_a \left(1-\cos(\Phi/f_a) \right)
$$
where $f_a$ is the axion decay constant and $m_a$ is the mass of the axion \cite{Guerra:2019srj,Delgado:2020udb}.

Certainly, there are many other possible choices of a potential term for the boson stars. In particular, there is a class of
flat potentials arising in the models with  gauge- and gravity-mediated supersymmetry breaking
mechanism \cite{Copeland:2009as,Hartmann:2012gw}. Such potentials may be of the logarithmic or the exponential
form, for example \cite{Copeland:2009as,Campanelli:2007um}
$$
U= \mu^2\eta^2\left[1-\exp\left(-\frac{\Phi^2}{\eta^2} \right)\right]
$$
where $\mu$ is the mass of the scalar field $\Phi$ and
the parameter $\eta$ is defines the mass scale below which supersymmetry is broken.

Domain of existence of the BSs is determined by the form of the potential. The maximal value $\omega_{max}$ corresponds to the
mass of the scalar excitations $\mu^2=\frac{d U}{d |\Phi|^2}$, the minimal value $\omega_{min}$ depends on explicit form of the
potential and  on the strength of the gravitational coupling $\alpha$. Hereafter we assume that $\mu=1$, without loss of
generality, hence in the EKG model $\omega_{max} = 1$. Since the Planck mass is defined as $M_{Pl}=1/\sqrt{G}$,
the EKG BSs can be interpreted as {\it macroscopic quantum states}, they are prevented from  gravitational collapse by the
uncertainty principle. The critical mass of the EKG BSs is $M\approx M_{Pl}^2/\mu$ \cite{Kaup:1968zz}, more massive BSs become
unstable w.r.t. linear fluctuations  \cite{Lee:1988av,Gleiser:1988rq}.
In the models with non-linear potentials, like \re{pot-quart}, \re{pot-six}, the BSs may have larger mass,
they represent lumps of a macroscopic self-gravitating Bose-Einstein condensate. In the discussion below we mainly focus on the
microscopic BSs in the EKG model and fix the value of the gravitational coupling $\alpha=0.5$.

\subsection{The ansatz and the field equations}

For the stationary spinning scalar field we can adopt a general  Ansatz with a harmonic time dependence:
\be
\label{scalans}
\Phi=f(r,\theta,\varphi)e^{-i(\omega t + n \varphi)},
\ee
where $r,\theta,\varphi$ are the usual spherical coordinates,
$\omega \geq 0$ is the angular frequency, $n \in {\mathbb{Z}}$
is the azimuthal winding number, and $f(r,\theta,\varphi)$ is a real spatial profile function.
Notably, harmonic time dependency of the scalar field does not affect the physical quantities, like
the stress-energy tensor \re{SET}. On the other hand, it allows us to evade scaling arguments of the Derrick's
theorem \cite{Derrick:1964ww},
which does not support existence of static scalar soliton solutions in three spatial dimensions.

Allowing an angular dependence for the profile function of the BSs requires considering a metric Ansatz
with sufficient generality. In particular, considering configurations with $n=0$, which carry no angular momentum, we
can make use of the line element without any spatial isometries
\be
\label{metrans}
ds^2=-F_0 dt^2 + F_1 dr^2 + F_2(rd\theta + S_1 dr )^2 + F_3(r\sin\theta d\varphi + S_2 dr +S_3 r d\theta)^2
\ee
where seven metric functions $F_0,F_1,F_2,F_3$ and $S_1,S_2,S_3$ depend on spherical coordinates $r,\theta,\varphi$
\cite{Herdeiro:2020kvf}.

By substituting the ansatz \re{scalans} into the scalar field equation \re{scaleq} we obtain
\begin{equation}
\begin{split}
&\frac{1}{\sqrt{-g}}\frac{\partial}{\partial r}\left(g^{rr}\sqrt{-g}\frac{\partial f}{\partial r} \right) +
\frac{1}{\sqrt{-g}}\frac{\partial}{\partial \theta}\left(g^{\theta\theta}\sqrt{-g}\frac{\partial f}{\partial \theta} \right) +
\frac{1}{\sqrt{-g}}\frac{\partial}{\partial \varphi}\left(g^{\varphi\varphi}\sqrt{-g}\frac{\partial f}{\partial r} \right) -\\
&-(n^2g^{\varphi\varphi} - 2 g^{\varphi t } + \omega^2 g^{tt}) f = \frac{d U}{d |\phi|^2 } f
\end{split}
\label{scalar-eqs}
\end{equation}

Note that on the spatial asymptotic the metric approaches the Minkowski spacetime, then the field equation \re{scalar-eqs}
tends to the usual Klein-Gordon equation with general solution for the scalar field
$f\sim \sum_{l,n} R_l(r)Y_{ln}(\theta,\varphi)$. Here
the radial part is
\be
R_l(r) \sim \frac{1}{\sqrt r } K_{l+\frac12}(r,\sqrt{\mu^2-\omega^2})
\ee
where $K_{l+\frac12}$ is the modified Bessel function of the first kind of order $l$ and $Y_{ln}(\theta,\varphi)$
are the real spherical harmonics, which form a complete basis on the sphere $S^2$ and integers $l\ge n$ are the usual
quantum numbers. Because of central character of gravitational interaction, this basis remains for any scalar multipole
configuration of the BSs. Furthermore, for each particular set of values of the
quantum numbers $l,n$, there are two types of the solutions, the parity even for even  $l$ and the parity-odd
for odd $l$. They are symmetric and anti-symmetric, respectively, under a reflection along the equatorial plane.
The spherical harmonics $Y_{ln}(\theta,\varphi)$ possess $2n$ $\varphi$-zeros, each describing a nodal longitude line and
$l-n$ $\theta$-zeros, each yielding a nodal latitude line. These nodal distributions define a multipolar configuration
of BSs  \cite{Herdeiro:2020kvf} briefly discussed below.

Simplest BSs are spherically symmetric \cite{Kaup:1968zz,Feinblum:1968nwc,Ruffini:1969qy}, in such a case $l = n = 0$ and
the profile function depends on the radial coordinate only,
$f=f(r)$. The corresponding line element can be reduced to the Schwarzschild type metric, it can be written as
\be
ds^2=-N(r)\sigma^2(r)dt^2 + \frac{dr^2}{N(r)} + r^2 (d\theta^2 + r^2 \sin^2 \theta~ d\varphi^2 )
\ee
with $N(r)=1-2m(r)/r$. Here $m(r)$ is so-called mass function, the Arnowitz-Deser-Misner (ADM) mass of the BS
is $M=\lim_{r\to \infty} m(r)$. Clearly, the angular momentum of such BS is zero.

\begin{figure}[h!]
\includegraphics[height=.30\textheight,  angle =-90]{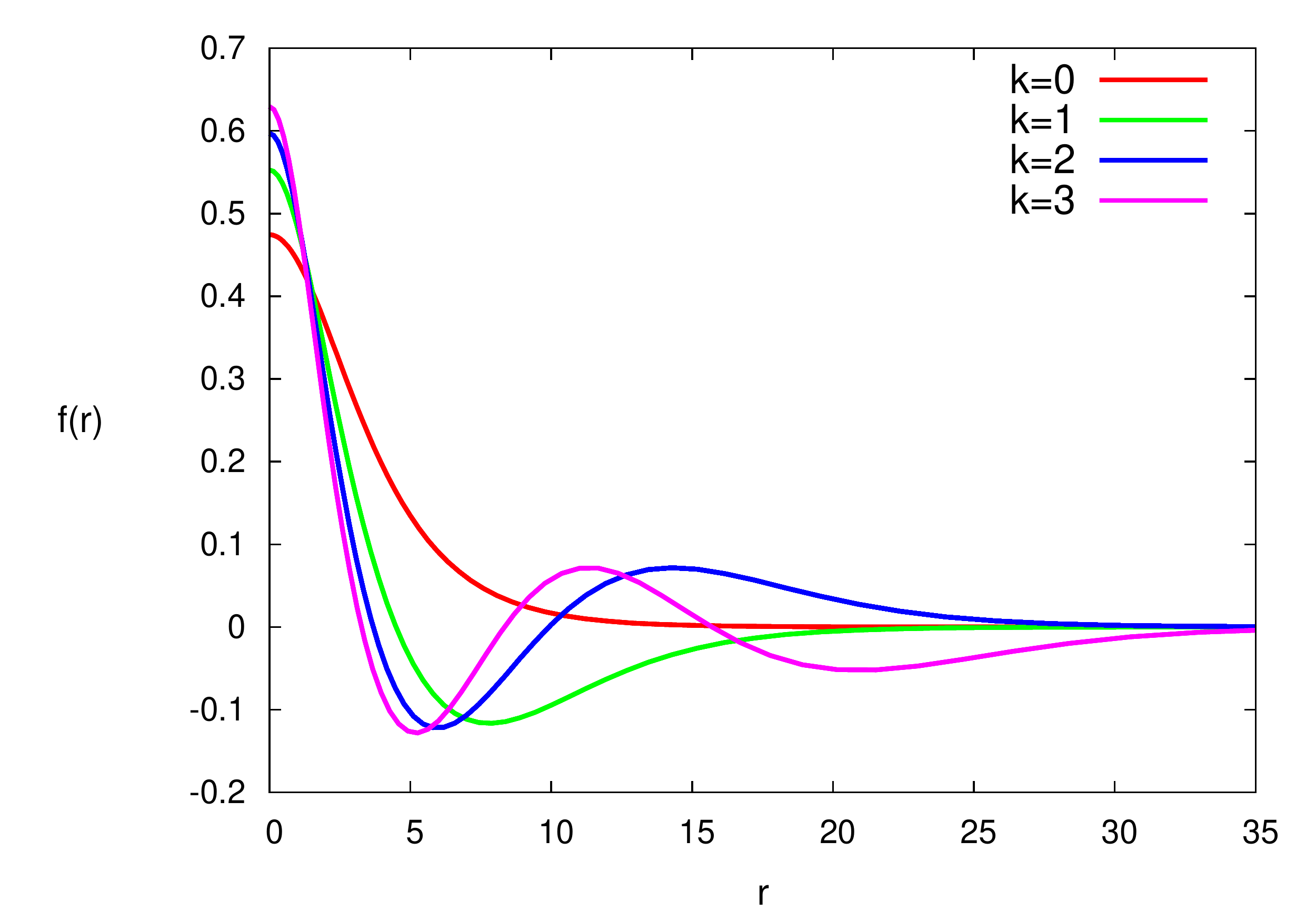}
\includegraphics[height=.30\textheight,  angle =-90]{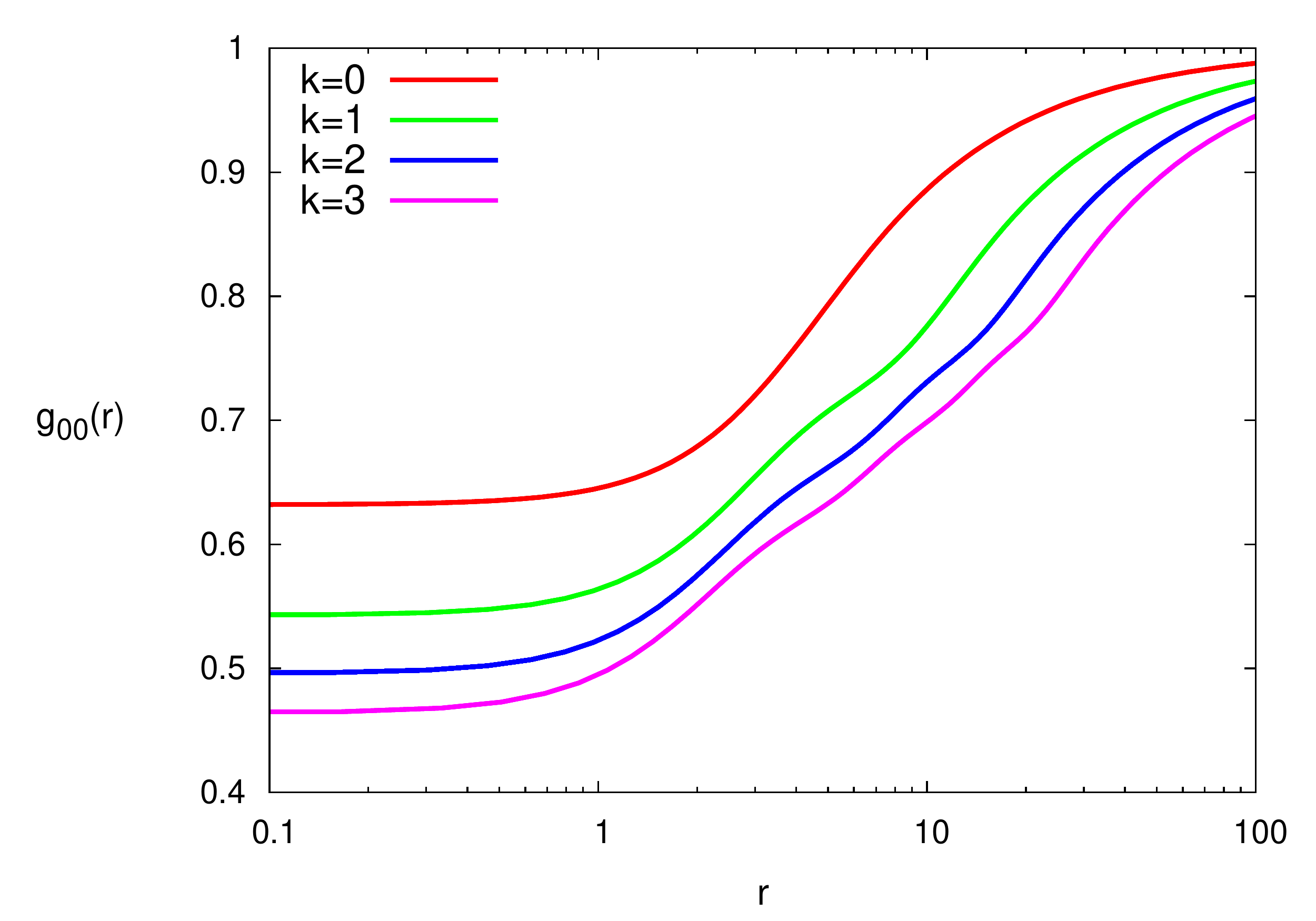}
\caption{\small
The profile functions of the scalar field (left) and
the metric component $g_{00}$ (right) of the non-rotating $n=0$ fundamental Einstein-Klein-Gordon boson star $k=0$
and its first three radial excitations are displayed on the first branch of solutions at $\omega=0.90$ as
functions of the radial coordinate.}
    \label{fig1}
\end{figure}

The resulting system of coupled ordinary differential equations on three radial functions
$f(r),\sigma(r)$ and $m(r)$ can be solved numerically, using, for example a shooting method \cite{Dias:2015nua}.
Along with the fundamental modeless mode, there is an infinite tower of radial excitations of the BSs
\cite{Jetzer:1991jr,Friedberg:1986tq},
they are classified according to the number of nodes $k$ of the scalar profile function  $f(r)$, see  Fig.~\ref{fig1}.

\begin{figure}[h!]
\includegraphics[height=.30\textheight,  angle =-90]{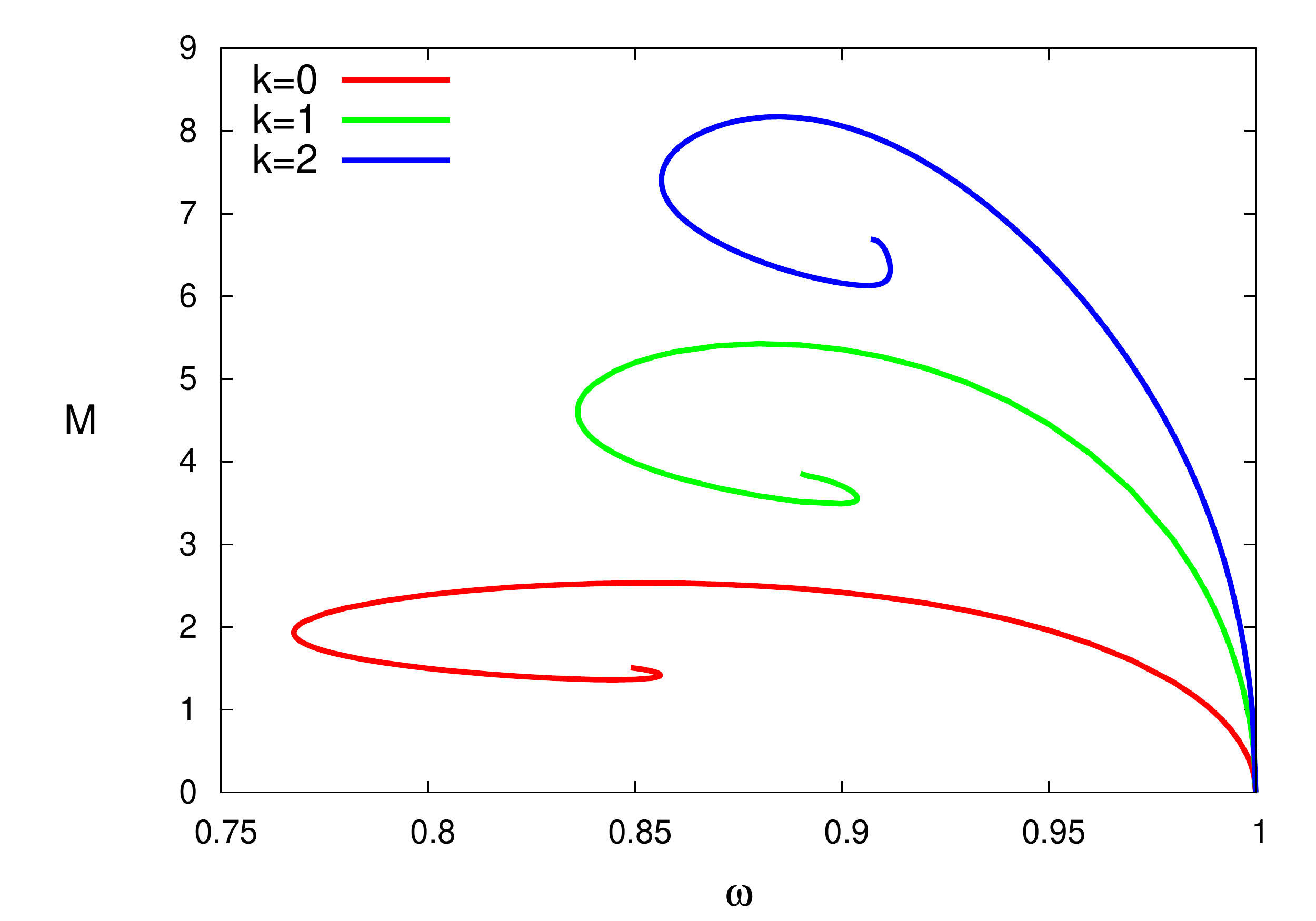}
\includegraphics[height=.30\textheight,  angle =-90]{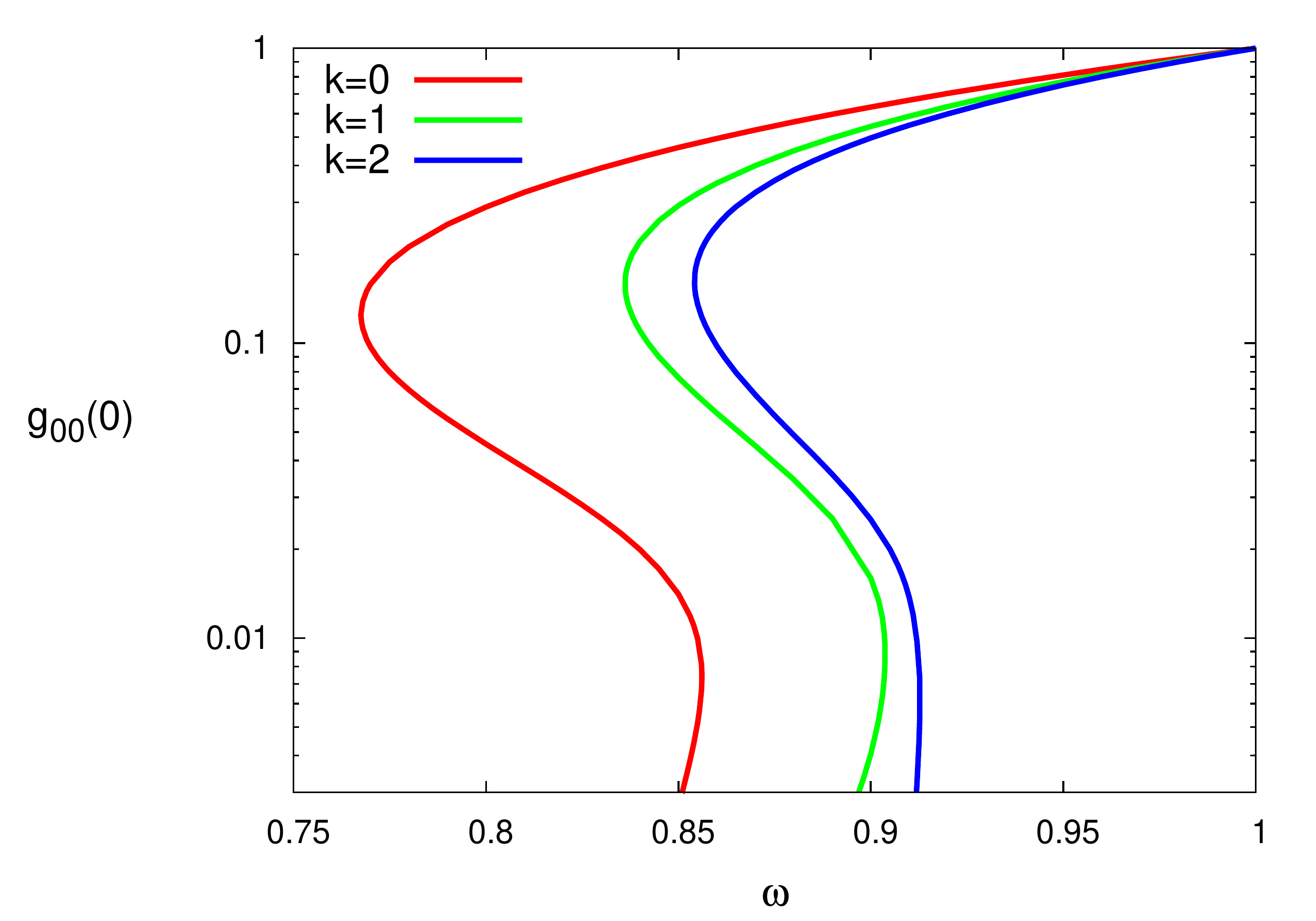}
\caption{\small Non-rotating $n=0$ fundamental ($k=0$) and radially excited
($k=1,2$) Einstein-Klein-Gordon boson stars.
The mass of the solutions (left plot) and
the minimal values of the metric component $g_{00}(0)$ (right plot)
are displayed as functions
of the angular frequency $\omega$.}
    \label{fig2}
\end{figure}

The fundamental nodeless ground state solution is an analog of the $1s$ hydrogen orbital. This branch of BSs emerges from the
vacuum fluctuations with angular part $Y_{00}$ at the maximal frequency $\omega_{max}$, given by the boson mass.
Notably, unlike the case of Q-balls in flat space,
where mass and charge diverge, these quantities vanish in this limit. Decreasing the frequency yields the
fundamental branch of solutions
which terminates at the first backbending of the curve, at which point it moves toward larger frequencies,
as seen in  Fig.~\ref{fig2}.
These solutions are stable with respect to linear
perturbations \cite{Seidel:1990jh}.
The curve then follows a spiraling/oscillating pattern, with successive backbendings, while
the minimum of the metric component $g_00(0)$ and the maximum of
the scalar profile function $f(0)$ show damped oscillations  \cite{Friedberg:1986tq}.
Both mass and charge tend to some finite limiting values at the centers of the corresponding spirals, see Fig.~\ref{fig2}.
Qualitatively, the appearance of the frequency-mass spiral may be related to oscillations in the alternating force balance between
the repulsive scalar interaction and the gravitational attraction in equilibria.
There is an infinite set of branches, leading towards a critical solution at the center of the spiral.
Plotting  the  $Q$  (instead of $M$) also yields similar curves.
The extremal values of the scalar field
profile function and the metric function $g_{00}$ at the center of the star
do not seem to be finite, with $f_{max}$ diverging and $g_{00}^{(min)}$ vanishing in this limit.

The radially excited spherically symmetric BSs also exhibit such spiraling behavior, as seen in Fig.~\ref{fig2}.
They emerge similarly from the vacuum at the maximal frequency.
These BSs posses higher mass, increase of the nodal number $k$ leads to increase of the
minimal critical frequency $\omega_{min}$, as seen in Fig.~\ref{fig2}, left plot.

\begin{figure}[h!]
\includegraphics[height=.30\textheight,  angle =-90]{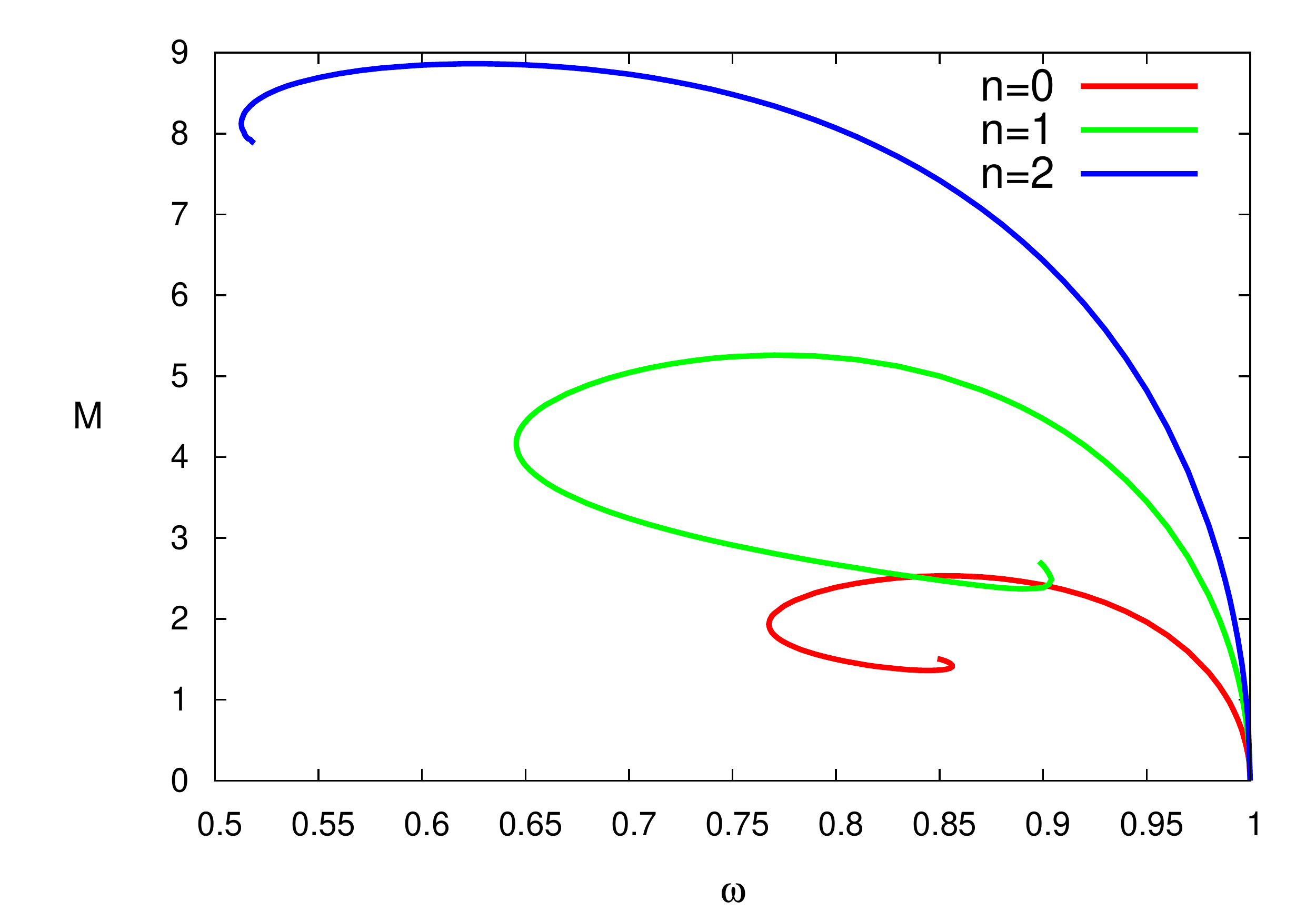}
\includegraphics[height=.30\textheight,  angle =-90]{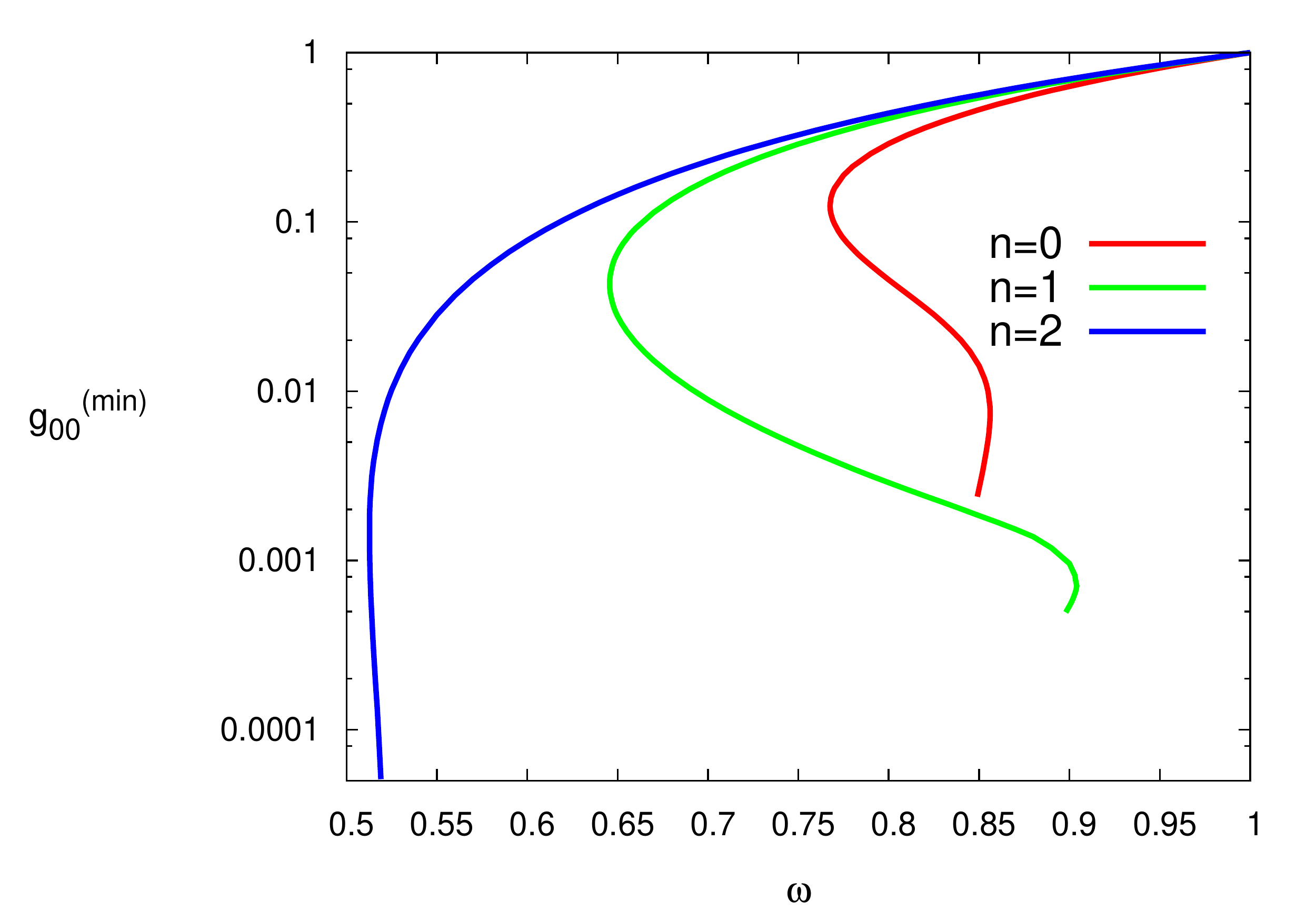}
\caption{\small Rotating Einstein-Klein-Gordon boson stars.
The mass of the solutions (left plot) and
the minimal  values of the metric component $g_{00}$ (right plot)
are displayed as functions
of the angular frequency $\omega$.}
    \label{fig3}
\end{figure}

Stationary spinning BSs are axially symmetric, their angular momentum is quantized in units of the azimuthal
winding number, $J=nQ$
\cite{Volkov:2002aj,Kleihaus:2005me,Kleihaus:2007vk}. Similar to the case of non-rotating spherically symmetric BSs,
they exhibit an analogous spiralling frequency/mass dependence. The rotating BSs exist in the EKG model
\cite{Silveira:1995dh,Schunck:1996he,Yoshida:1997qf,Grandclement:2014msa} and in the model with solitonic potential \re{pot-six}
\cite{Kleihaus:2005me,Kleihaus:2007vk} as well as in other systems.
The mass and the charge of the rotating BSs  are much higher than
the fundamental spherically symmetric counterparts, as seen in Fig.~\ref{fig3}. The energy density distribution of
these solutions is torus-like, the scalar field is vanishing at the origin, it possess a maximal value in the equatorial
plane.

Remarkably, rapidly rotating BSs develop an ergoregion where the Killing vector field  $\xi=\partial_t$
becomes spacelike \cite{Kleihaus:2007vk,Grandclement:2014msa}, or equally, $g_{tt}< 0$. Topologically, this region
represent a torus. The existence of ergoregions is typical
for a Kerr black hole, for the BSs it is an indication of instability of the configuration.
The instability mechanism is related to the rotational superradiance \cite{Brito:2015oca}, an excited relativistic
BSs decays into less energetic state via emission of scalar quanta and gravitational waves. On the other hand,
for the Kerr black hole with synchronized scalar hair \cite{Hod:2012px,Herdeiro:2014goa} the superradiance mechanism
may induce transitions from the $n=0$ state to higher $n$ solutions \cite{Delgado:2019prc}.

The frequency dependence of rotating nodeless axially symmetric BSs
is similar to that of the fundamental $n=0$ solutions,
the mass (and the angular momentum) form a spiral, as $\omega$ varies,
while the minimum of the metric component $g_{00}$
and the maximum of the scalar function $f$ shows damped oscillations, see Fig.~\ref{fig3}.
The minimal value of the angular frequency $\omega_{min}$ is decreasing as the winding number $n$ increases. Further,
for each value of integer winding number $n$,
there are two types of spinning BSs possessing
different parity, so called parity-even and parity-odd rotating hairy BHs
\cite{Yoshida:1997qf,Volkov:2002aj,Kleihaus:2005me,Kleihaus:2007vk,Radu:2008pp}.
These configurations are symmetric or anti-symmetric, respectively, with respect to a
reflection through the equatorial plane, i.e. under $\theta \to \pi -\theta$. In other words, the scalar field of the
parity-odd BSs posses an angular node at $\theta = \pi/2$.

\begin{figure}[h!]
\includegraphics[height=3.6cm,width=0.95\textwidth]{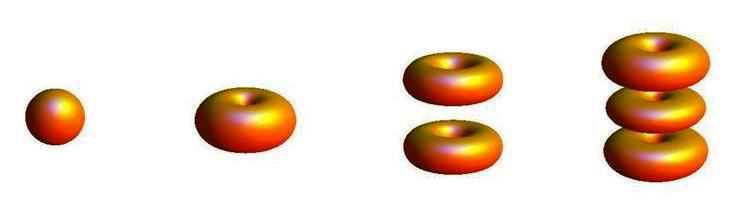}
\caption{\small Einstein-Klein-Gordon boson stars. Surfaces of constant energy density of
the (i) fundamental $n=0$ solution; (ii) parity-even $n=1$ rotating boson star;
(iii) parity-odd $n=1$ rotating boson star; and (iv) angularly excited parity-even $n=1$ boson star, from left to right, all
configurations at $\omega = 0.92$ on the first branch at $\alpha=0.5$.}
    \label{fig4}
\end{figure}

The energy density distribution
of rotating BSs with positive parity forms a torus, while the energy density of rotating
parity-odd BSs corresponds to a double torus, see see Fig.~\ref{fig4}. More generally, there is a sequence of
angularly excited BSs with some number of nodes of the scalar field in $\theta$-direction \cite{Brihaye:2008cg},
which are closely related to
the real spherical harmonics $Y_{lm}(\theta,\varphi)$. For example, the angular part of the
$n=1$ spinning parity-even BSs corresponds to the
harmonic $Y_{11}$ while the angular part of the corresponding parity-odd BSs corresponds to the harmonic $Y_{21}$, the triple torus
configuration, displayed in the right plot of Fig.~\ref{fig4}, corresponds to the harmonic $Y_{31}$, etc.

\begin{figure}[h!]
\includegraphics[height=.30\textheight,  angle =-90]{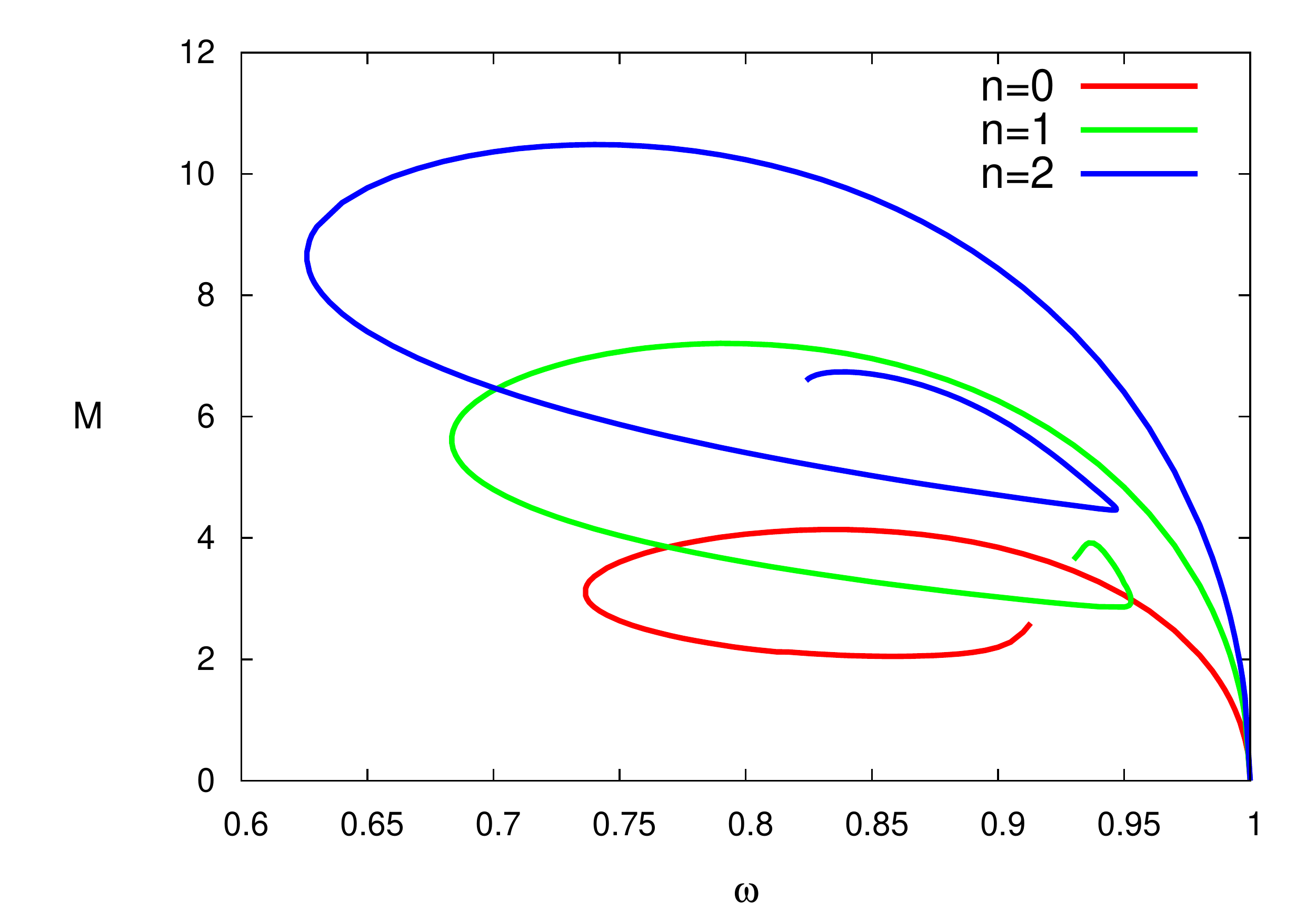}
\includegraphics[height=.30\textheight,  angle =-90]{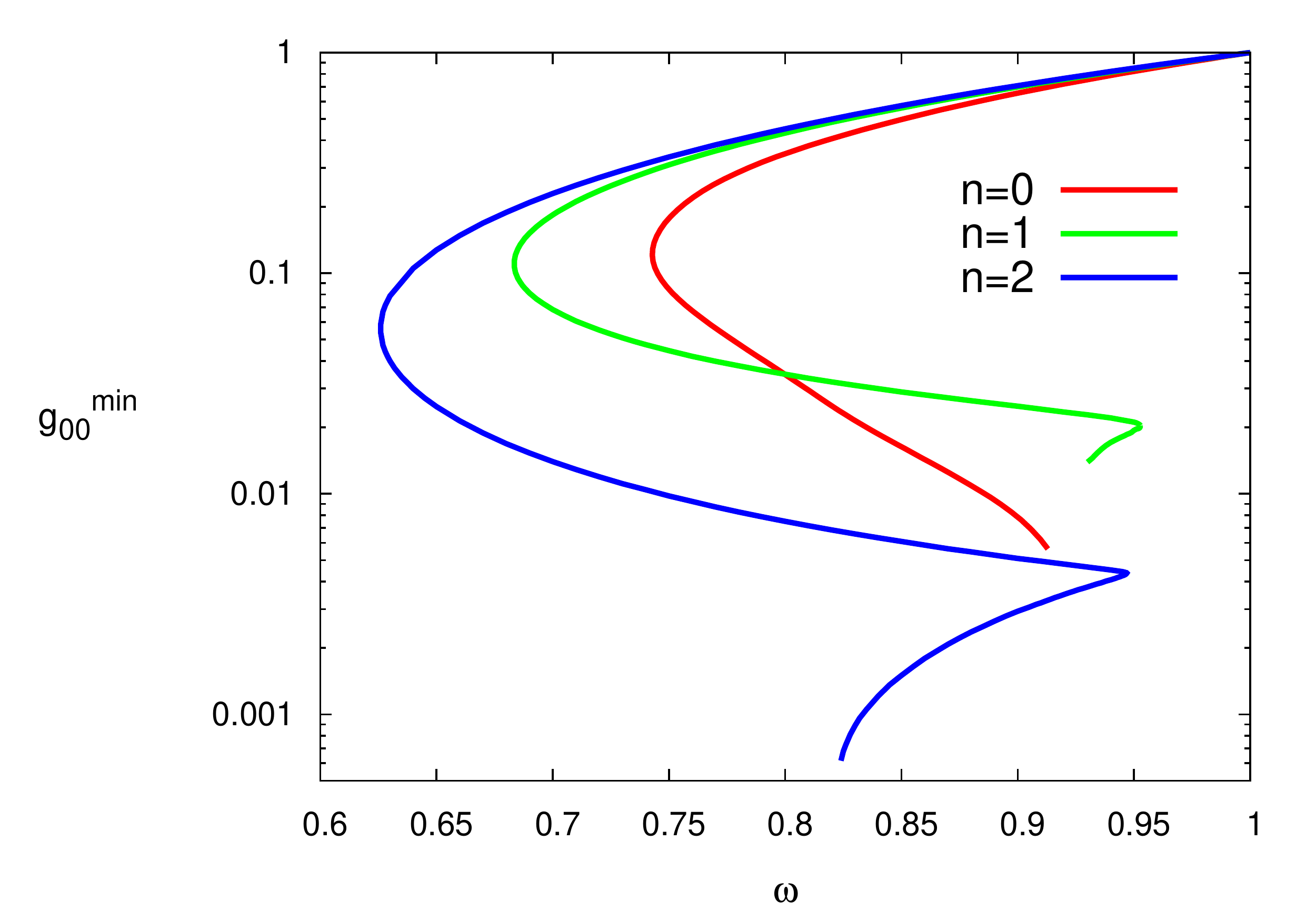}
\caption{\small Rotating parity-odd Einstein-Klein-Gordon boson stars.
The mass of the solutions (left plot) and
the minimal  values of the metric component $g_{00}$ (right plot)
are displayed as functions
of the angular frequency $\omega$.}
    \label{fig5}
\end{figure}

The mass and the charge of both parity-even and parity-odd EKG BSs exhibit similar
spiraling behavior, cf Figs.~\ref{fig3},\ref{fig5}.
However, the situation becomes different for the rotating radially excited axially symmetric BSs with non-zero angular momentum
\cite{Collodel:2017biu}.
In such a case, there are two branches of solutions, merging and ending at the minimal values of the charge and the
mass of the configurations, the second branch extends all the way back to the upper critical value of the
frequency $\omega_{max}$, forming a loop.

Notably, the gravitational interaction stabilizes the parity-odd BSs even in the limit $n=0$.
This axially-symmetric configuration with zero angular momentum represents a pair of boson stars, a saddle point solution
of the EKG model \cite{Yoshida:1997nd,Palenzuela:2006wp,Herdeiro:2021mol,Herdeiro:2020kvf}. Its existence is related to
a delicate force balance between the repulsive scalar interaction and gravity. Indeed, is the flat space Q-balls
are in phase, they attract each other, if they are out of phase, there is a
repulsive scalar force between them \cite{Battye:2000qj,Bowcock:2008dn}. The
inversion of the sign of the scalar field function $\Phi$ under reflections
$\theta \to \pi - \theta$ corresponds to the shift of the phase $\omega \to \omega +\pi$. Hence,
the static pair of BSs with a single node of the scalar field on the symmetry axis, can be
thought of as the limit of negative parity spinning configurations considered in \cite{Kleihaus:2007vk}.

The curves of the mass/frequency dependency of the pair of BSs are different from the case of
a single spherical BS \cite{Herdeiro:2021mol}. Instead of the paradigmatic spiraling curve one finds
a truncated scenario with only two branches, ending at a limiting solution
with finite values of ADM mass and Noether charge.

\begin{figure}
\begin{center}
\includegraphics[width=0.95\textwidth]{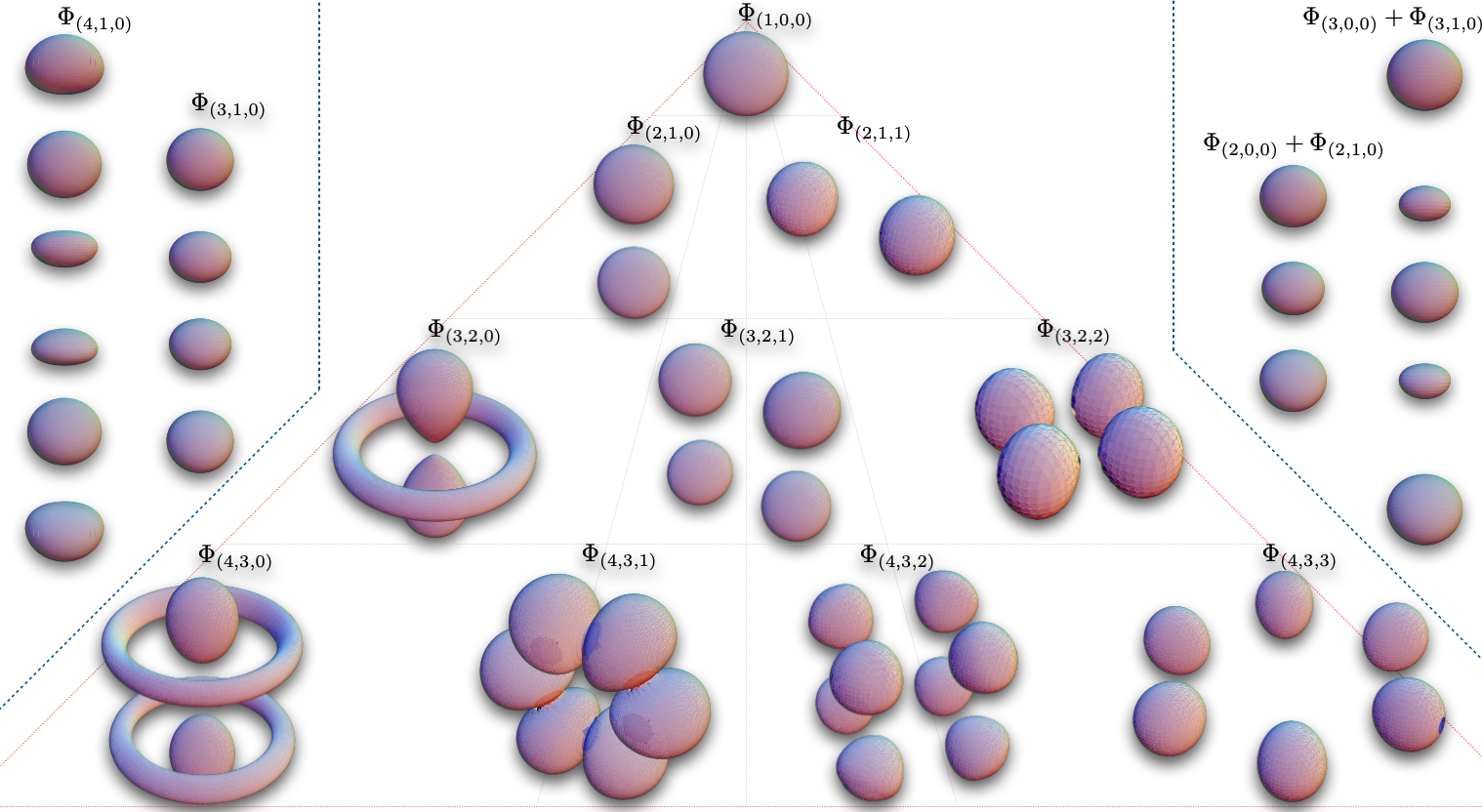}
\caption{\small Surfaces of constant energy density for a selection of multicomponent BSs in the EKG model.
Reprinted (without modification)
from \cite{Herdeiro:2020kvf}. $\copyright$ 2021 The Authors of
\cite{Herdeiro:2020kvf}, under the CC BY 4.0 license.}
\label{fig6}
\end{center}
\end{figure}

Furthermore, scalar repulsion can be
balanced by the gravitational attraction in various multicomponent bounded systems of BSs \cite{Herdeiro:2021mol,Herdeiro:2020kvf}.
Fig.~\ref{fig6} displays an overview of a selection of multipolar EKG BSs with various structure of nodes \cite{Herdeiro:2020kvf}.
Constructing these solutions we do not impose any restrictions of symmetry, they all arise as corresponding linearized
perturbations of the scalar field in the asymptotic region, as $\omega$ approaches the mass threshold. Gravitational attraction
stabilizes the excitations with nodal structure of the $\Phi(r,\theta,\varphi)\sim R_k(r)Y_{ln}(\theta,\varphi))$ wavefunctions.
As the angular frequency decreases, the mass and the charge of the multicomponent configurations increase, however, the nodal structure
remains unaffected \cite{Herdeiro:2020kvf}. Similar to the fundamental spherically symmetric solution,
the fundamental branch of the multicomponent BSs ends in a spiraling/oscillating pattern.
Clearly, all these solutions do not exist in Minkowsky space-time.

\begin{figure}[tb]
    \begin{center}
   \includegraphics[height=5.cm,width=0.99\textwidth]{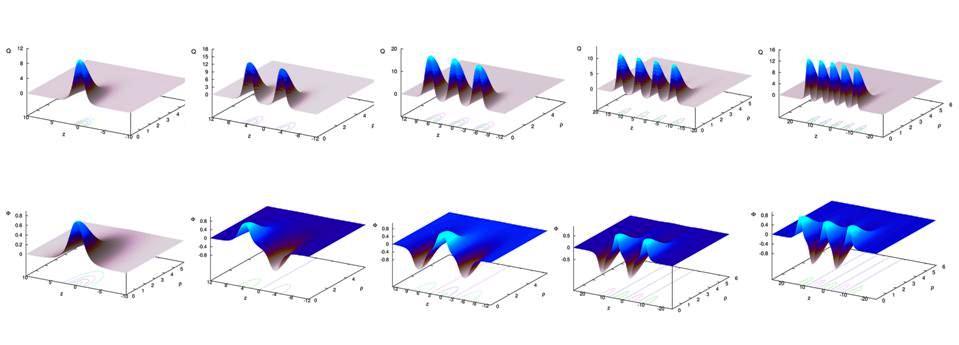}
    \end{center}
    \caption{\small
Chains of BSs with one to five constituents
on the  first  branch for $\alpha=0.25$ at $\omega/\mu=0.80$:
$3d$ plots of the $U(1)$ scalar charge distributions (upper row) and
the scalar field $\Phi$ (bottom row)
versus the coordinates $\rho = r\sin \theta$ and $z=r\cos\theta$.}
   \label{fig7}
\end{figure}

Analogous multipolar configurations with zero angular momentum
exist in models with various potentials. For example, the chains of BSs
in the system with sixtic potential \re{pot-six} were discussed in \cite{Herdeiro:2021mol}.
Fig.~\ref{fig7} exhibits a few examples of these chains.

In such a case the pattern of dynamical evolution of the multicomponent BSs becomes different from the above-discussed
EKG systems. Chains with an odd number of constituents
show a spiraling behavior for their  mass and charge in terms of their angular frequency, similarly to a
single fundamental BS,
as long as the gravitational coupling is relatively small.
For larger coupling however, the spiral is replaced
by a lace with two ends approaching the mass threshold,
each branch corresponding to the dominance of either of the two states, and with a self-crossing.
In other words, the branch of odd chains
bifurcates with the fundamental branch of radially excited spherical boson stars.

For the even chains we do not observe the endless spiraling scenario, on the second, or on the third branch
the configuration evolves toward a limiting solution which retain basically two central constituents,
whose metric function $g_{00}$ exhibits two sharp peaks,
reaching a very small value, while the scalar field features
two sharp opposite extrema located right at the location of
these peaks \cite{Herdeiro:2021mol}.

\section*{Acknowledgements}
I am very grateful to Carlos Herdeiro, Jutta Kunz, Victor Loiko,
Ilya Perapechka and Eugen Radu for valuable collaboration, many
results of our joint work are reviewed in this brief survey. I
gratefully acknowledge the networking support by the COST Actions
CA16104. This work was supported in part by the Ministry of
Science and High Education of Russian Federation, project
FEWF-2020-0003 and by the BLTP JINR Heisenberg-Landau program
2020.


\begin{thebibliography}{99.}%
\bibitem{Jetzer:1991jr}
P.~Jetzer,
Phys. Rept. \textbf{220} (1992), 163-227
\bibitem{Liddle:1992fmk}
A.~R.~Liddle and M.~S.~Madsen,
Int. J. Mod. Phys. D \textbf{1} (1992), 101-144
\bibitem{Schunck:2003kk}
F.~E.~Schunck and E.~W.~Mielke,
Class. Quant. Grav. \textbf{20} (2003), R301-R356
\bibitem{Liebling:2012fv}
S.~L.~Liebling and C.~Palenzuela,
Living Rev. Rel. \textbf{15} (2012), 6
\bibitem{Wheeler:1955zz}
J.~A.~Wheeler,
Phys. Rev. \textbf{97} (1955), 511-536
\bibitem{Manton:2004tk}
N.~S.~Manton and P.~Sutcliffe, \textit{Topological solitons}.
 ( Cambridge University
Press, Cambridge, 2004)
\bibitem{Shnir2018} Y.M.~Shnir, {\it Topological and Non-Topological Solitons in Scalar Field Theories}.
 ( Cambridge University
Press, Cambridge, 2018)
\bibitem{Rosen}G.~Rosen,
J.\ Math.\ Phys.\ {\bf 9} (1968) 996, 999
\bibitem{Friedberg:1976me}
  R.~Friedberg, T.~D.~Lee and A.~Sirlin,
  Phys.\ Rev.\ D {\bf 13} (1976) 2739
\bibitem{Coleman:1985ki}
  S.~R.~Coleman,
  Nucl.\ Phys.\ B {\bf 262} (1985) 263
   Erratum: [Nucl.\ Phys.\ B {\bf 269} (1986) 744].
\bibitem{Kaup:1968zz}
  D.~J.~Kaup,
  Phys.\ Rev.\  {\bf 172} (1968) 1331.
\bibitem{Feinblum:1968nwc}
D.~A.~Feinblum and W.~A.~McKinley,
Phys. Rev. \textbf{168} (1968) no.5, 1445
\bibitem{Ruffini:1969qy}
  R.~Ruffini and S.~Bonazzola,
  Phys.\ Rev.\  {\bf 187} (1969) 1767.
\bibitem{Kleihaus:2005me}
  B.~Kleihaus, J.~Kunz and M.~List,
  Phys.\ Rev.\ D {\bf 72} (2005) 064002
\bibitem{Kleihaus:2007vk}
  B.~Kleihaus,  J.~Kunz, M.~List and I.~Schaffer,
  Phys.\ Rev.\ D {\bf 77} (2008) 064025
\bibitem{Friedberg:1986tq}
R.~Friedberg, T.~D.~Lee and Y.~Pang,
Phys. Rev. D \textbf{35} (1987), 3658
\bibitem{Seidel:1990jh}
E.~Seidel and W.~M.~Suen,
Phys. Rev. D \textbf{42} (1990), 384-403
\bibitem{Colpi:1986ye}
M.~Colpi, S.~L.~Shapiro and I.~Wasserman,
Phys. Rev. Lett. \textbf{57} (1986), 2485-2488
\bibitem{Herdeiro:2015tia}
C.~A.~R.~Herdeiro, E.~Radu and H.~R\'unarsson,
Phys. Rev. D \textbf{92} (2015) no.8, 084059
\bibitem{Sanchis-Gual:2021phr}
N.~Sanchis-Gual, C.~Herdeiro and E.~Radu,
[arXiv:2110.03000 [gr-qc]].
\bibitem{Suarez:2013iw}
A.~Su\'arez, V.~H.~Robles and T.~Matos,
Astrophys. Space Sci. Proc. \textbf{38} (2014), 107-142
\bibitem{Lee:1986ts}
T.~D.~Lee,
Phys. Rev. D \textbf{35} (1987), 3637
\bibitem{Hawley:2000dt}
S.~H.~Hawley and M.~W.~Choptuik,
Phys. Rev. D \textbf{62} (2000), 104024
\bibitem{Hui:2016ltb}
L.~Hui, J.~P.~Ostriker, S.~Tremaine and E.~Witten,
Phys. Rev. D \textbf{95} (2017) no.4, 043541
\bibitem{Guerra:2019srj}
D.~Guerra, C.~F.~B.~Macedo and P.~Pani,
JCAP \textbf{09} (2019) no.09, 061
\bibitem{Delgado:2020udb}
J.~F.~M.~Delgado, C.~A.~R.~Herdeiro and E.~Radu,
JCAP \textbf{06} (2020), 037
\bibitem{Copeland:2009as}
E.~J.~Copeland and M.~I.~Tsumagari,
Phys. Rev. D \textbf{80} (2009), 025016
\bibitem{Hartmann:2012gw}
B.~Hartmann and J.~Riedel,
Phys. Rev. D \textbf{87} (2013) no.4, 044003
\bibitem{Campanelli:2007um}
L.~Campanelli and M.~Ruggieri,
Phys. Rev. D \textbf{77} (2008), 043504
\bibitem{Cardoso:2019rvt}
V.~Cardoso and P.~Pani,
Living Rev. Rel. \textbf{22} (2019) no.1, 4
\bibitem{Glampedakis:2017cgd}
K.~Glampedakis and G.~Pappas,
Phys. Rev. D \textbf{97} (2018) no.4, 041502
\bibitem{Herdeiro:2021lwl}
C.~A.~R.~Herdeiro, A.~M.~Pombo, E.~Radu, P.~V.~P.~Cunha and
N.~Sanchis-Gual,
JCAP \textbf{04} (2021), 051
\bibitem{Palenzuela:2007dm}
C.~Palenzuela, L.~Lehner and S.~L.~Liebling,
Phys. Rev. D \textbf{77} (2008), 044036
\bibitem{Bezares:2017mzk}
M.~Bezares, C.~Palenzuela and C.~Bona,
Phys. Rev. D \textbf{95} (2017) no.12, 124005
\bibitem{Palenzuela:2017kcg}
C.~Palenzuela, P.~Pani, M.~Bezares, V.~Cardoso, L.~Lehner and
S.~Liebling,
Phys. Rev. D \textbf{96} (2017) no.10, 104058
\bibitem{Lee:1988ag}
K.~M.~Lee, J.~A.~Stein-Schabes, R.~Watkins and L.~M.~Widrow,
Phys. Rev. D \textbf{39} (1989), 1665
\bibitem{Lee:1991bn}
  C.~H.~Lee and S.~U.~Yoon,
  Mod.\ Phys.\ Lett.\ A {\bf 6} (1991) 1479.
\bibitem{Kusenko:1997vi}
  A.~Kusenko, M.~E.~Shaposhnikov and P.~G.~Tinyakov,
  Pisma Zh.\ Eksp.\ Teor.\ Fiz.\  {\bf 67} (1998) 229
   [JETP Lett.\  {\bf 67} (1998) 247]
\bibitem{Anagnostopoulos:2001dh}
 K.~N.~Anagnostopoulos,
   M.~Axenides, E.~G.~Floratos and N.~Tetradis,
  Phys.\ Rev.\ D {\bf 64} (2001) 125006
\bibitem{Gulamov:2015fya}
  I.~E.~Gulamov et al
  Phys.\ Rev.\ D {\bf 92} (2015) no.4,  045011
\bibitem{Gulamov:2013cra}
  I.~E.~Gulamov, E.~Y.~Nugaev and M.~N.~Smolyakov,
  Phys.\ Rev.\ D {\bf 89} (2014) no.8,  085006
\bibitem{Panin:2016ooo}
  A.~G.~Panin and M.~N.~Smolyakov,
  Phys.\ Rev.\ D {\bf 95} (2017) no.6,  065006
\bibitem{Nugaev:2019vru}
E.~Y.~Nugaev and A.~V.~Shkerin,
J. Exp. Theor. Phys. \textbf{130} (2020) no.2, 301-320
\bibitem{Loiko:2019gwk}
V.~Loiko and Y.~Shnir,
Phys. Lett. B \textbf{797} (2019), 134810
\bibitem{Jetzer:1989av}
P.~Jetzer and J.~J.~van der Bij,
Phys. Lett. B \textbf{227} (1989), 341-346
\bibitem{Jetzer:1989us}
P.~Jetzer,
Phys. Lett. B \textbf{231} (1989), 433-438
\bibitem{Jetzer:1992tog}
P.~Jetzer, P.~Liljenberg and B.~S.~Skagerstam,
Astropart. Phys. \textbf{1} (1993), 429-448
\bibitem{Lee:1988av}
T.~D.~Lee and Y.~Pang,
Nucl. Phys. B \textbf{315} (1989), 477
\bibitem{Gleiser:1988rq}
M.~Gleiser,
Phys. Rev. D \textbf{38} (1988), 2376 [erratum: Phys. Rev. D
\textbf{39} (1989) no.4, 1257]
\bibitem{Pugliese:2013gsa}
D.~Pugliese, H.~Quevedo, J.~A.~Rueda H. and R.~Ruffini,
Phys. Rev. D \textbf{88} (2013), 024053
\bibitem{Kleihaus:2009kr}
B.~Kleihaus, J.~Kunz, C.~Lammerzahl and M.~List,
Phys. Lett. B \textbf{675} (2009), 102-115
\bibitem{Kumar:2014kna}
S.~Kumar, U.~Kulshreshtha and D.~Shankar Kulshreshtha,
Class. Quant. Grav. \textbf{31} (2014), 167001
\bibitem{Astefanesei:2003qy}
D.~Astefanesei and E.~Radu,
Nucl. Phys. B \textbf{665} (2003), 594-622
\bibitem{Kichakova:2013sza}
O.~Kichakova, J.~Kunz and E.~Radu,
Phys. Lett. B \textbf{728} (2014), 328-335
\bibitem{wald}
 R.~M. Wald,
 {\it General Relativity},
 (University of Chicago Press, Chicago, 1984).
\bibitem{Levin:2010gp}
  A.~Levin and V.~Rubakov,
  Mod.\ Phys.\ Lett.\ A {\bf 26} (2011) 409.
\bibitem{Loiko:2018mhb}
  V.~Loiko, I.~Perapechka and Y.~Shnir,
  Phys.\ Rev.\ D {\bf 98} (2018) no.4,  045018
\bibitem{Friedberg:1986tp}
R.~Friedberg, T.~D.~Lee and Y.~Pang,
Phys. Rev. D \textbf{35} (1987), 3640
\bibitem{Bernal:2009zy}
A.~Bernal, J.~Barranco, D.~Alic and C.~Palenzuela,
Phys. Rev. D \textbf{81} (2010), 044031
\bibitem{Gleiser:1988ih}
M.~Gleiser and R.~Watkins,
Nucl. Phys. B \textbf{319} (1989), 733-746
\bibitem{Kain:2021rmk}
B.~Kain,
Phys. Rev. D \textbf{103} (2021) no.12, 123003
\bibitem{Schunck:1996he}
F.~E.~Schunck and E.~W.~Mielke,
Phys. Lett. A \textbf{249} (1998), 389-394
\bibitem{Cardoso:2007az}
V.~Cardoso, P.~Pani, M.~Cadoni and M.~Cavaglia,
Phys. Rev. D \textbf{77} (2008), 124044
\bibitem{Collodel:2017biu}
L.~G.~Collodel, B.~Kleihaus and J.~Kunz,
Phys. Rev. D \textbf{96} (2017) no.8, 084066
\bibitem{Silveira:1995dh}
V.~Silveira and C.~M.~G.~de Sousa,
Phys. Rev. D \textbf{52} (1995), 5724-5728
\bibitem{Yoshida:1997qf}
S.~Yoshida and Y.~Eriguchi,
Phys. Rev. D \textbf{56} (1997), 762-771
\bibitem{Grandclement:2014msa}
P.~Grandclement, C.~Som\'e and E.~Gourgoulhon,
Phys. Rev. D \textbf{90} (2014) no.2, 024068
\bibitem{Volkov:2002aj}M.S.~Volkov and E.~Wohnert,
Phys.\ Rev.\  D  {\bf 66} (2002)  085003.
\bibitem{Radu:2008pp}E.~Radu and M.S.~Volkov,
Phys.\ Rept.\  {\bf 468}  (2008)  101.
\bibitem{Battye:2000qj}
R.~Battye and P.~Sutcliffe,
Nucl. Phys. B \textbf{590} (2000), 329-363
\bibitem{Bowcock:2008dn}
P.~Bowcock, D.~Foster and P.~Sutcliffe,
J. Phys. A \textbf{42} (2009), 085403
\bibitem{Yoshida:1997nd}
S.~Yoshida and Y.~Eriguchi,
Phys. Rev. D \textbf{55} (1997), 1994-2001
\bibitem{Palenzuela:2006wp}
C.~Palenzuela, I.~Olabarrieta, L.~Lehner and S.~L.~Liebling,
Phys. Rev. D \textbf{75} (2007), 064005
\bibitem{Herdeiro:2021mol}
C.~A.~R.~Herdeiro, J.~Kunz, I.~Perapechka, E.~Radu and Y.~Shnir,
Phys. Rev. D \textbf{103} (2021) no.6, 065009
\bibitem{Herdeiro:2020kvf}
C.~A.~R.~Herdeiro, J.~Kunz, I.~Perapechka, E.~Radu and Y.~Shnir,
Phys. Lett. B \textbf{812} (2021), 136027
\bibitem{Deppert:1979au}
W.~Deppert and E.~W.~Mielke,
Phys. Rev. D \textbf{20} (1979), 1303-1312
\bibitem{Mielke:1980sa}
E.~W.~Mielke and R.~Scherzer,
Phys. Rev. D \textbf{24} (1981), 2111
\bibitem{Derrick:1964ww}
  G.~H.~Derrick,
  J.\ Math.\ Phys.\  {\bf 5} (1964) 1252.
\bibitem{Dias:2015nua}
\'O.~J.~C.~Dias, J.~E.~Santos and B.~Way,
Class. Quant. Grav. \textbf{33} (2016) no.13, 133001
\bibitem{Brito:2015oca}
R.~Brito, V.~Cardoso and P.~Pani,
Lect. Notes Phys. \textbf{906} (2015), pp.1-237
\bibitem{Hod:2012px}
  S.~Hod,
  Phys.\ Rev.\ D {\bf 86} (2012) 104026
  Erratum: [Phys.\ Rev.\ D {\bf 86} (2012) 129902]
\bibitem{Herdeiro:2014goa}
  C.~A.~R.~Herdeiro and E.~Radu,
  Phys.\ Rev.\ Lett.\  {\bf 112} (2014) 221101
\bibitem{Delgado:2019prc}
J.~F.~M.~Delgado, C.~A.~R.~Herdeiro and E.~Radu,
Phys. Lett. B \textbf{792} (2019), 436-444
\bibitem{Brihaye:2008cg}
Y.~Brihaye and B.~Hartmann,
Phys. Rev. D \textbf{79} (2009), 064013

\end{thebibliography}
\end{document}